\documentstyle[aps,prc,preprint,epsfig,feynmf,floats]{revtex}

\newcommand{\feynbox}[2]{\makebox[#1]{\parbox{#1}{\makebox[#1]{#2}}}}

\newcommand{\dl}[2]{\fmf{dashes}{#1,#2} \fmfdot{#1} \fmfdot{#2}}
\newcommand{\cs}[2]{\fmf{dashes,left}{#1,#2} \fmfdot{#1} \fmfdot{#2}}
\newcommand{\cd}[2]{\fmf{dashes,right}{#1,#2} \fmfdot{#1} \fmfdot{#2}}
\newcommand{\cb}[2]{\fmf{plain,left}{#1,#2} \fmf{plain,right}{#1,#2}
			\fmfdot{#1} \fmfdot{#2}}

\newcommand{\dLo}[3]{\fmf{phantom,label=$#3$}{#1,#2}}

\newcommand{\labellinesingle}[4]{\fmflabel{$#3$}{#1} \fmflabel{$#4$}{#2}}
\newcommand{\labelline}[7]{\fmflabel{$#5$}{#1} \fmflabel{$#7$}{#4}
	\fmf{phantom,label=$#6$}{#2,#3}}

\newcommand{\gengr}[3]{
\feynbox{#1 \unitlength}{
  \begin{fmfgraph}(#1,#2)
    \fmfstraight
    \fmfleft{k1,k2} 
    \fmfright{p1,p2}  
    \fmf{plain}{k1,s11,p1}
    \fmf{plain}{k2,t11,p2}    \fmffreeze
    \fmf{phantom}{k1,s21,s22,p1}
    \fmf{phantom}{k2,t21,t22,p2}    \fmffreeze
    \fmf{phantom}{k1,s31,s32,s33,p1}
    \fmf{phantom}{k2,t31,t32,t33,p2}    \fmffreeze
    \fmf{phantom}{k1,s41,s42,s43,s44,p1}
    \fmf{phantom}{k2,t41,t42,t43,t44,p2}    \fmffreeze
    #3
  \end{fmfgraph}}}
\newcommand{\gengrL}[3]{
\feynbox{#1 \unitlength}{
  \begin{fmfgraph*}(#1,#2)
    \fmfstraight
    \fmfleft{k1,k2} 
    \fmfright{p1,p2}  
    \fmf{plain}{k1,s11,p1}
    \fmf{plain}{k2,t11,p2}    \fmffreeze
    \fmf{phantom}{k1,s21,s22,p1}
    \fmf{phantom}{k2,t21,t22,p2}    \fmffreeze
    \fmf{phantom}{k1,s31,s32,s33,p1}
    \fmf{phantom}{k2,t31,t32,t33,p2}    \fmffreeze
    \fmf{phantom}{k1,s41,s42,s43,s44,p1}
    \fmf{phantom}{k2,t41,t42,t43,t44,p2}    \fmffreeze
    #3
  \end{fmfgraph*}}}

\newcommand{\itOPEgraphOne}[2]  {\gengr{#1}{#2}{\dl{s41}{t42} \dl{s43}{t44}}}
\newcommand{\itOPEgraphTwo}[2]  {\gengr{#1}{#2}{\dl{s41}{t42} \dl{s44}{t43}}}
\newcommand{\itOPEgraphThree}[2]{\gengr{#1}{#2}{\dl{s42}{t41} \dl{s43}{t44}}}
\newcommand{\itOPEgraphFour}[2] {\gengr{#1}{#2}{\dl{s42}{t41} \dl{s44}{t43}}}
\newcommand{\sbgraphOne}[2]     {\gengr{#1}{#2}{\dl{s41}{t43} \dl{s42}{t44}}}
\newcommand{\sbgraphTwo}[2]     {\gengr{#1}{#2}{\dl{s44}{t42} \dl{s43}{t41}}}

\newcommand{\opegraphLOne}[5]    {\gengrL{#1}{#2}{
	\dl{s21}{t22}	\dLo{s11}{t11}{#5}
	\labellinesingle{k1}{p1}#3
	\labellinesingle{k2}{p2}#4}}
\newcommand{\opegraphLTwo}[5]    {\gengrL{#1}{#2}{
	\dl{s22}{t21}	\dLo{s11}{t11}{#5}
	\labellinesingle{k1}{p1}#3
	\labellinesingle{k2}{p2}#4}}
\newcommand{\sbgraphLOne}[6]{\gengrL{#1}{#2}{
	\dl{t43}{s41}	\dLo{t42}{s42}{#5}
	\dl{s42}{t44}	\dLo{s43}{t43}{#6}
	\labelline{k1}{s41}{s42}{p1}#3
	\labelline{k1}{s41}{s42}{p1}#3
	\labelline{k2}{t44}{t43}{p2}#4}}
\newcommand{\sbgraphLTwo}[6]{\gengrL{#1}{#2}{
	\dl{t42}{s44}	\dLo{s43}{t43}{#5}
	\dl{s43}{t41}	\dLo{t42}{s42}{#6}
	\labelline{k1}{s43}{s44}{p1}#3
	\labelline{k2}{t42}{t41}{p2}#4}}

\newcommand{\FeynOBE}[2]          {\gengr{#1}{#2}{\dl{s11}{t11}}}
\newcommand{\FeynLaddergraph}[2]  {\gengr{#1}{#2}{\dl{s21}{t21} \dl{s22}{t22}}}
\newcommand{\FeynTBEX}[2]         {\gengr{#1}{#2}{\dl{s21}{t22} \dl{s22}{t21}}}
\newcommand{\FeynTBEvertone}[2]   {\gengr{#1}{#2}{\dl{s11}{t11} \cs{t31}{t33}}}
\newcommand{\FeynTBEverttwo}[2]   {\gengr{#1}{#2}{\dl{s11}{t11} \cd{s31}{s33}}}
\newcommand{\FeynTBEmr}[2]        {\gengr{#1}{#2}{
				\fmf{phantom}{s11,u1,u2,t11}
				\fmffreeze
				\dl{s11}{u1} \dl{t11}{u2} \cb{u1}{u2}}}

\fmfset{thin}{0.5pt}
\fmfset{dash_len}{2mm}
\fmfset{dot_size}{2pt}

\begin{document}
\begin{fmffile}{fmfriy}
\draft

\tightenlines

\preprint{NT@UW-99-52}

\title{Restoration of rotational invariance of bound states on the light front}
\author{Jason R.~Cooke, Gerald A.~Miller, and Daniel R.~Phillips}
\address{
Department of Physics \\
University of Washington \\
Box 351560 \\
Seattle WA 98195-1560, USA}
\date{\today{}}
\maketitle

\begin{abstract}
We study bound states in a model with scalar nucleons interacting via an
exchanged scalar meson using the Hamiltonian formalism on the light front. In
this approach manifest rotational invariance is broken when the Fock space is
truncated.  By considering an effective Hamiltonian that takes into account
two meson exchanges, we find that this breaking of rotational invariance is
decreased from that which occurs when only one meson exchange is included.
The best improvement occurs when the states are weakly bound.
\end{abstract}

\pacs{PACS number(s): 
21.45.+v, 
03.65.Ge, 
03.65.Pm, 
11.10.Ef  
}

\section{Introduction}

Covariant field theory can be coerced into giving solutions to non-perturbative
problems such as bound states via the Bethe-Salpeter equation.  There are
technical difficulties in arriving at those solutions, but they are manifestly
covariant.  An alternative approach that also provides covariant results is
obtained by using a Hamiltonian theory and the corresponding Schr\"odinger
equation to solve for the bound states.  A consequence of this covariance is
that the commutation relations between the operators that constitute the
four-momentum ($P^\mu$) and the four-angular momentum ($M^{\mu\nu}$) give a
representation of the Poincar\'e algebra.  There are many different
formulations of Hamiltonian theories, corresponding to different forms
of dynamics, as discussed by Dirac \cite{Dirac:1949cp}.  The components
of the $P^\mu$ and $M^{\mu\nu}$ operators can be classified as either
dynamic (dependent on the interaction) or kinematic.  Different forms of
dynamics result in different separations of $P^\mu$ and $M^{\mu\nu}$
into dynamic operators and kinematic operators.  The two forms that we
will be concerned with are the equal-time and light-front formalisms.
The equal-time form gives four dynamical operators, while the
light-front form yields only three dynamical ones. Having a small
number of dynamical operators is good since this simplifies some
calculations.

Using the canonical coordinate system, with $x^\mu = (x^0,x^1,x^2,x^3)$
and defining the commutation relations at equal time ($x^0=t$), we
obtain the most familiar Hamiltonian theory, an equal-time theory.  For
this system, rotations and translations are kinematical, while the
Hamiltonian and generators of boosts are dynamical. The fact that the
generators of boosts are dynamical implies that when the full
Hamiltonian is truncated, it will no longer transform appropriately
under boosts.  This implies that the equal-time Hamiltonian may not be
well-suited for relativistic problems. 

An alternative coordinate system is obtained by using light-front variables
$x^\pm = x^0 \pm x^3$ to construct $x^\mu = (x^+,x^-,x^1,x^2)$.  High-energy
experiments are naturally described using these coordinates, rather than
equal-time ones.  This is because the front of a beam traveling at the
speed of light in
the (negative) three-direction is defined by a surface where $x^+$ is
constant.  The description of the interaction of this beam with a target
is much simpler if described in terms of light-front variables
\cite{Brodsky:1997de,Miller:1997cr} than if an equal-time variables are
used.  As an example, the Bjorken $x$ variable used to describe deep
inelastic scattering experiments is simply the ratio of the plus
momentum of the struck constituent particle to the total plus momentum
of the bound state.

Using the light-front coordinate system and defining the commutation relations
at equal light-front time ($x^+=t_{\text{LF}}$), we obtain a
light-front Hamiltonian
theory \cite{Brodsky:1997de,Harindranath:1996hq,Heinzl:1998kz}.
For this choice, rotation
about the three axis, translations in the one, two, and plus directions, and
the generators of boosts in the one, two, and plus directions are
kinematical, while the
Hamiltonian $(P^-)$ and rotations about the one and two directions are
dynamical.  Since the boosts are kinematical, the light-front Hamiltonian is
useful for relativistic problems, such as strongly-bound systems, even when
the Hamiltonian is truncated.  However, since two of the rotations are
dynamical, the momentum operator four-vector will not be covariant under
those rotations when the Hamiltonian is truncated.  This lack of manifest
rotational invariance presents some difficulties for bound-state calculations.

A key benefit of using the light front is that the vacuum can be very
simple.  For our theory, the masses of the particles are large.  All
massive particles and anti-particles have positive plus momentum, and
the total plus momentum is a conserved quantity.  Thus, there is no
condensation, and the vacuum (with $p^+=0$) is empty, the Fock space
vacuum.  Thus, the Hilbert space for this theory is the Fock space.
Diagrams that couple to the vacuum vanish, so the number of
non-trivial light-front time-ordered diagrams is greatly reduced
compared to the equal-time theory.

\subsection{Difficulties with the light front}

An untruncated light-front Hamiltonian will commute with the total relative
angular momentum operator, since the total momentum commutes with
the relative momentum.  Thus, for the scalar theory we consider here,
the eigenstates of the full Hamiltonian will also be eigenstates of the angular
momentum.  However, as mentioned earlier, a Fock-space truncation of the
light-front  Hamiltonian results in the momentum operator four-vector losing
covariance under rotations.  Hence $J^2$ and the truncated Hamiltonian do not
commute and this implies that the eigenstates of the truncated Hamiltonian
will not be eigenstates of the angular momentum.

How will this violation of rotational invariance affect physical observables?
A way to observe this violation is to note that on the
light front, rotational invariance about the three-axis is maintained.  This
allows us to classify states as eigenstates of $J_3$ with eigenvalues $m$.  We
compare the energies of states with the same ``relative angular momentum''
quantum number $l$ but different  $m$ values.
We will define the ``relative angular momentum'' operator in
section \ref{uxsec}.  If the Hamiltonian were
rotationally invariant, the energies should be the same; the breaking of
rotational invariance causes the energies to be different
\cite{Trittmann:1997ga}.

We expect that the higher Fock-space components of the full Hamiltonian will
be small if the coupling constant is small enough.  Thus, truncation at
successively higher orders should reduce the violation of rotational
invariance of the truncated Hamiltonian.  In particular, by retaining enough
terms in the perturbation expansion of the Hamiltonian, the violation of
rotational invariance can be reduced to an arbitrarily small amount, provided
only that a perturbation expansion is valid.  Thus, the only real question is:
How many terms are required?

\subsection{Outline of the rest of the paper}

The purpose of this paper is to explore how truncation at different orders in
the effective potential  affects the breaking of rotational invariance of the
spectra.  The model we are using is a Lagrangian with neutral scalar particles
coupled via a $\phi^2 \chi$ interaction to other neutral scalar particles.  The
Bethe-Salpeter equation for this Lagrangian is discussed, and the ladder
approximation to the Bethe-Salpeter equation and its spectra are reviewed.  We
also derive the light-front Hamiltonian from that Lagrangian and write out the
light-front Schr\"odinger equation for bound states consisting of two
particles.  An approximation to the light-front potential that causes the
light-front Schr\"odinger equation to be equivalent to the ladder
Bethe-Salpeter equation is discussed.  This approximation will be called the
{\em uncrossed approximation}.  The one-boson-exchange (OBE) effective
potential and two-boson-exchange (TBE) effective potential are derived in this
approximation.  All of this is discussed in more detail in section
\ref{ourmodel}.

Following the development of the model, we look at some numerical calculations
in section \ref{results}.  For our tests, we look at states composed of two
massive scalar particles bound by the effects of the exchange of a lighter
scalar particle.  We calculate the spectra (the coupling constant versus
bound-state mass curves) using the Bethe-Salpeter equation in the ladder
approximation.  Since the Bethe-Salpeter equation is manifestly covariant, its
solutions are states with  definite angular momentum.  We also calculate the
spectra using a light-front Schr\"odinger equation with the OBE and the 
OBE+TBE potentials.  Because of the lack of manifest rotational invariance,
these wavefunctions do not have definite angular momentum.  An artificial
``relative angular momentum'' operator is constructed, and  
a partial-wave
decomposition is performed on these wavefunctions to analyze the various
angular momentum states present in them.  The problem of trying to classify
these states is discussed.  We then plot the spectra for the wavefunctions
classified as the lowest-lying $p$- and $d$-wave states for the
Bethe-Salpeter, OBE Schr\"odinger, and the OBE+TBE Schr\"odinger equations.

In section \ref{conclusions}, we summarize our findings: the higher-order
terms help to partially restore the breaking of rotational invariance
introduced by using only the OBE potential.  This restoration is largest when
the states are weakly bound.

We stress that the point of this paper is {\em not} to solve the Lagrangian of
this model exactly.  Instead, we seek to understand how rotational invariance
is broken by the light-front Hamiltonian in the uncrossed approximation, how
it can be partially restored by keeping higher-order graphs, and also to
compare the results from the two truncated Hamiltonians to the results of the
``exact'' theory (the ladder Bethe-Salpeter equation).  Discussion of the
inclusion of the crossed-graph contributions will be given in a forthcoming
paper \cite{Cooke}. 

This study is closely connected to the work of Schoonderwoerd, Bakker, and
Karmanov \cite{Schoonderwoerd:1998pk}, who considered two-body scattering in
perturbation theory using a model similar to ours.  Off-shell scattering
amplitudes were computed at order $g^4$ and it was found that for low momenta,
the TBE diagrams contributed much less than the iterated OBE diagrams.  This
lead to a conjecture that the lightly-bound states should be well described by
the OBE potential, with the TBE potential playing a minor role.  These results
encourage us to calculate and compare bound states incorporating the OBE and
OBE+TBE diagrams.  We note that some of the difficulties present in 
\cite{Schoonderwoerd:1998pk} are avoided here since there are no singularities
in the bound-state integral equation.

While this manuscript was in preparation, a paper by Sales {\it et.\ al.} 
\cite{Sales:1999ec} appeared.  In that work, a model similar to ours was used
to consider bound states.  Their study included the same two truncations that
we use, as well as the a comparison to the Bethe-Salpeter equation results.
However, the investigation of Ref.\ \cite{Sales:1999ec} focused on the ground
state, while here we are concerned with the rotational invariance of the
excited states.

\section{Our model} \label{ourmodel}

We consider two distinguishable uncharged scalars $\phi=(\phi_1,\phi_2)$ with
mass $M$ (which we will refer to as {\em nucleons}), that couple to a third,
uncharged scalar $\chi$ with mass $\mu$ (which we will refer to as a
{\em meson}) by a $\phi^2 \chi$ interaction.  This $\phi^2 \chi$ theory, which
can be considered a massive extension of the Wick-Cutkosky model
\cite{WickCutkosky}, has been used by Schoonderwoerd, Bakker,
and Karmanov \cite{Schoonderwoerd:1998pk} on the light front to study
scattering states.  The Lagrangian is
\begin{eqnarray}
\mathcal{L} &=& 
\frac{1}{2} \left( \partial_\mu \phi \partial^\mu \phi -   M^2 \phi^2 \right) +
\frac{1}{2} \left( \partial_\mu \chi \partial^\mu \chi - \mu^2 \chi^2 \right) +
\frac{g M}{2} \chi \phi^2. \label{thelagrangian}
\end{eqnarray}
This Lagrangian will be used in two formalisms, the Bethe-Salpeter equation and
the Hamiltonian approach.

\subsection{Bethe-Salpeter equation}

The Bethe-Salpeter equation \cite{Nambu:1950rs,Schwinger:1951ex,%
Schwinger:1951hq,Gell-Mann:1951rw,Salpeter:1951sz} provides a way of
describing bound states based on Feynman propagators and kernels constructed
from covariant quantum field theory, and thus is manifestly covariant.  The
equation for the bound state of nucleons 1 and 2 can be written as
\begin{eqnarray}
G K \psi &=& \psi, \label{bse}
\end{eqnarray}
$G$ is the free two-particle propagator,
which is the product of two one-particle propagators, $G=S_1 S_2$, $\psi$ is
the four-dimensional Bethe-Salpeter amplitude, and $K$ is the sum of all two-particle
irreducible two-to-two Feynman graphs,
\begin{eqnarray}
K &=&
\FeynOBE{30}{30} +
\FeynTBEX{30}{30} +
\FeynTBEvertone{30}{30} +
\FeynTBEverttwo{30}{30} +
\FeynTBEmr{30}{30} +
\cdots \, .
\end{eqnarray}
There are no nucleon exchange graphs since we treat only the case where the
two nucleons are distinguishable.

We consider the ladder approximation to the Bethe-Salpeter equation,
\begin{eqnarray}
g^2 G K_{\text{OBE}} \psi &=& \psi, \label{lbseforget}
\end{eqnarray}
which is obtained by replacing $K$ in Eq.\ (\ref{bse}) with
$g^2 K_{\text{OBE}}$, the graph due to one-boson-exchange,
\begin{eqnarray}
K_{\text{OBE}} &=& \frac{1}{g^2} \FeynOBE{30}{20}.
\end{eqnarray}
This definition of $K_{\text{OBE}}$ makes it
independent of the coupling constant.
Making this approximation leaves the Bethe-Salpeter equation covariant.  This
implies that the Bethe-Salpeter amplitudes $\psi$ have definite angular
momentum $l$, and 
hence the energies of the states are degenerate for different $m$ projections
of the same angular momentum.

This equation can be simplified in the center-of-momentum frame.  In
that frame, once the total energy is defined as $P^0$,
the four-momentum of the second particle 
is given by $k_2^\mu = P^\mu - k_1^\mu$, and thus the Bethe-Salpeter equation
is effectively a one-particle equation.  The remainder of the discussion in
this section will be done in the c.m.\ frame.  We then can rewrite Eq.\ 
(\ref{lbseforget}), taking into account the explicit symmetries of the
equation, as
\begin{eqnarray}
\left[ g_{\text{LBSE}}^{n,l,m}(E) \right]^2 G(E) K_{\text{OBE}}(E) \psi_{n,l,m}
&=& \psi_{n,l,m}, \label{lbse}
\end{eqnarray}
where $E$ is an arbitrary energy, $\psi_{n,l,m}$ is the $n^{\text{th}}$
Bethe-Salpeter amplitude with angular momentum $l$ and three-projection $m$, and 
$g_{\text{LBSE}}^{n,l,m}(E)$ is the coupling constant which yields that
bound-state Bethe-Salpeter amplitude with $E$ as the bound-state energy.
We call this $g(E)$ the spectrum of the ladder
Bethe-Salpeter equation for the corresponding Bethe-Salpeter amplitude.  The Greens
function $G(E)$ and the OBE kernel $K_{\text{OBE}}(E)$ are functions
of the energy in the c.m.\ frame and are implicitly effective
one-particle operators.

The comparison we draw is only between the ladder Bethe-Salpeter equation and
the light-front Hamiltonian that corresponds to that approximation. 
The exact nature of the correspondence is discussed in section \ref{uxsec}.
 Since we
are not looking at the solutions to the full theory, for our purposes it does
not matter that there are sizeable
differences between the solutions to the full Bethe-Salpeter equation and the
ladder approximation when the coupling constant is large 
\cite{LevineWright1,LevineWright2,LevineWright3,Nieuwenhuis:1996mc}.

\subsection{Light-front Hamiltonian}

To obtain the light-front Hamiltonian from the Lagrangian in Eq.\ 
(\ref{thelagrangian}), we follow the approach used by Miller
\cite{Miller:1997cr} and many others (see the review \cite{Brodsky:1997de})
to write the light-front Hamiltonian ($P^-$) as the sum of a free,
non-interacting part and a term containing the interactions. This is
accomplished by using the energy-momentum tensor in
\begin{eqnarray}
P^\mu={1\over2}\int dx^-d^2x_\perp\;T^{+\mu}(x^+=0,x^-,\bbox{x}_\perp).
\end {eqnarray}
The usual relations determine  $T^{+\mu}$, with
\begin{eqnarray}
T^{\mu\nu}=-g^{\mu\nu}{\cal L} +\sum_r{\partial{\cal L}\over\partial
  (\partial_\mu\phi_r)}\partial^\nu\phi_r,
\label{tmunu}
\end{eqnarray}
in which the degrees of freedom are labeled by $\phi_r$.

It is worthwhile to consider the limit in which the interactions between the 
fields are removed.  This will allow us to define the free Hamiltonian $P^-_0$
and to display  the necessary commutation relations.  The energy-momentum
tensor of the non-interacting fields is defined as $T_0^{\mu\nu}$.  Use of
Eq.\ (\ref{tmunu})  leads to the result
\begin{eqnarray}
T^{\mu\nu}_0 &=&
\partial^\mu \phi \partial^\nu \phi - \frac{g^{\mu\nu}}{2}
\left[\partial_\sigma \phi \partial^\sigma \phi-M^2\phi^2\right] 
+ \partial^\mu \chi \partial^\nu \chi - \frac{g^{\mu\nu}}{2}
\left[\partial_\sigma \chi \partial^\sigma \chi-\mu^2\chi^2\right],
\end{eqnarray}
with
\begin{eqnarray}
T^{+-}_0 
= \bbox{\nabla}_\perp \phi \cdot \bbox{\nabla}_\perp\phi + M^2 \phi^2 +
\bbox{\nabla}_\perp \chi \cdot \bbox{\nabla}_\perp\chi + \mu^2 \chi^2.
\end{eqnarray}

The scalar nucleon fields can be expressed in terms of creation and
destruction operators:
\begin{eqnarray}
\phi_i(x)&=&
\int \frac{d^2 k_\perp dk^+ \, \theta(k^+)}{(2\pi)^{3/2}\sqrt{2k^+}} \left[
a_i(\bbox{k})e^{-ik\cdot x} +a_i^\dagger(\bbox{k})e^{ik\cdot x}\right],
\end{eqnarray}
where $i=1,2$ is a particle index,
$k\cdot x={1\over2}(k^-x^++k^+x^-)-\bbox{k}_\perp \cdot \bbox{x}_\perp$ with
$k^-= \frac{M^2 + \bbox{k}_\perp^2}{k^+}$,
and $\bbox{k}\equiv(k^+,\bbox{k}_\perp)$.
Note that $k^-$ is such that the particles are on the mass shell, which
is a consequence of using a Hamiltonian theory.
The $\theta$ function restricts $k^+$ to positive values.
Likewise, the scalar meson field is given by
\begin{eqnarray}
\chi(x)&=&
\int \frac{d^2 k_\perp dk^+ \, \theta(k^+)}{(2\pi)^{3/2}\sqrt{2k^+}} \left[
a_\chi(\bbox{k})e^{-ik\cdot x} +a_\chi^\dagger(\bbox{k})e^{ik\cdot x}\right],
\end{eqnarray}
where $k^-= \frac{\mu^2 + \bbox{k}_\perp^2}{k^+}$,
so that the mesons are also on the
mass shell.  The non-vanishing commutation relations  are
\begin{eqnarray}
\left[a_\alpha(\bbox{k}),a_\alpha^\dagger(\bbox{k}')\right] &=&
\delta(\bbox{k}_\perp-\bbox{k}'_\perp)
\delta(k^+-k'^+), \label{comm}
\end{eqnarray}
where $\alpha = 1,2,\chi$ is a particle index.
The commutation relations are defined at equal light-front
time, $x^+=0$.

The derivatives appearing in the quantity $T^{+-}_0$ are evaluated and then one
sets $x^+$ to 0 to obtain the result
\begin{eqnarray}
P^-_0 &=& 
\int_k \, \left[
\frac{M^2 + \bbox{k}_\perp^2}{k^+}
\left( a_1^\dagger(k) a_1(k) + a_2^\dagger(k) a_2(k) \right) 
+ \frac{\mu^2 + \bbox{k}_\perp^2}{k^+}
a_\chi^\dagger(k) a_\chi(k)
\right], \label{p0minusop}
\end{eqnarray}
with $\int_k = \int d^2 k_\perp dk^+ \, \theta(k^+)$.  Eq.\ (\ref{p0minusop})
has the interpretation of an operator that counts the light-front
energy $k^-$ (which is $\frac{M^2 + \bbox{k}_\perp^2}{k^+}$ for the 
nucleons and $\frac{\mu^2 + \bbox{k}_\perp^2}{k^+}$ for the mesons)
of all of the particles.

We now consider the interacting part of the Lagrangian, ${\mathcal L}_I$.
An analysis similar to that for the non-interacting parts
yields the interacting part of the light-front
Hamiltonian $P_I^-$;
\begin{eqnarray}
P_I^- &=& \sum_{i=1,2} \frac{gM}{2} \int_k \int_{k'}
\frac{1}{(2\pi)^{3/2} \sqrt{2 k^+ k'^+(k^++k'^+)} } \nonumber\\ 
&& \phantom{\sum_{i=1,2} \frac{gM}{2} \int_k \int_{k'}} \times \left\{
\left[
2 a_i^\dagger(k+k') a_\chi(k') a_i(k)
+ a_\chi^\dagger(k+k') a_i(k') a_i(k) 
\right] \right. \nonumber\\
&& \phantom{\sum_{i=1,2} \frac{gM}{2} \int_k \int_{k'} \times \left\{
\right.}\left.
+ \mbox{Hermitian conjugate} \right\}.
\label{hamint}
\end{eqnarray}
For our theory, the Hilbert space is the Fock space.  Eq.~(\ref{hamint})
is self-adjoint on that Hilbert space, since the equation is written in
terms of the Fock operators.
The total light-front Hamiltonian is given by $P^- = P^-_0 + P^-_I$.

\subsection{Hamiltonian bound-state equations}
\label{uxsec}

We will be studying the bound states of two distinguishable nucleons in the
light-front analogue of old-fashioned (time-ordered) perturbation theory:
light-front time-ordered perturbation theory (LFTOPT).  Using this
perturbation theory for our Hamiltonian, we can write out a two-nucleon
effective Hamiltonian ``eigenvalue'' equation (a light-front Schr\"odinger
equation) where the potential is expanded in powers of the coupling constant
$g$.  This is not a true eigenvalue equation since the potential is
energy dependent.
We write this in the form of Eq.\ (\ref{lbse})
\begin{eqnarray}
\left[P_0^- +V \Big(g^n_{\text{LFSE}}(P^-),P^- \Big) \right]|\psi^n\rangle
&=& |\psi^n\rangle P^- \label{fullse} \\
V( g,P^- ) &=& \sum_{i=1}^\infty g^{2i} V_{(2i)}(P^-), \label{fullpot}
\end{eqnarray}
where $P^-$ is an arbitrary light-front energy, $|\psi^n\rangle$ is the 
the $n^{\text{th}}$ wavefunction, and $g^n_{\text{LFSE}}(P^-)$ is the 
coupling constant which yields that bound-state wavefunction with $P^-$ as
the bound-state energy.  We call this $g(P^-)$ the spectrum of the 
light-front Schr\"odinger equation for the corresponding wavefunction.
Only even powers of $g$ appear in $V$ since every meson emitted must be
absorbed.  Also, since $V_{(2i)}$ is the potential due to the exchange
of $i$ mesons, we call $V_{(2)}=V_{\text{OBE}}$ and $V_{(4)}=V_{\text{TBE}}$.
The potentials $V_{(2i)}$ can be calculated from the light-front
time-ordered diagrams.

We want to approximate our potential $V$ so that Eq.\ (\ref{fullse}) is
physically equivalent to the ladder Bethe-Salpeter equation.  This
approximation of $V$ will be called the {\em uncrossed approximation}.
By physically equivalent, we mean that the spectra of the potential $V$
should reproduce the spectrum for the states of the Bethe-Salpeter
equation, excluding the so-called ``abnormal'' states \cite{WickCutkosky}.
It is well known how to reduce the Bethe-Salpeter equation to a
physically equivalent Hamiltonian
(Schr\"odinger-type) equation.  For an extensive discussion of this issue of
defining the potential equivalent to a sum of Feynman graphs in the
equal-time case see, for instance, Klein \cite{klein}, Phillips and Wallace
\cite{Phillips:1996eb}, Lahiff and Afnan \cite{Lahiff:1997bj}, and for
examples on the light front, Ligterink and Bakker \cite{Ligterink:1995tm}.
The general procedure to get the effective potential due to $n$ boson exchange
takes two steps.  First, write the sum of all Feynman graphs obtained from
iteration of the Bethe-Salpeter kernel with $n$ boson exchanges.  Then, write
that sum in terms of LFTOPT graphs, and discard all graphs which are not
two-particle-irreducible with respect to the light-front two-body propagator,
\begin{eqnarray}
G_{\text{LF}}\left(P^-\right) &=& \frac{1}{P^- - P^-_0}.
\end{eqnarray}
The graphs which remain after this procedure constitute the $V_{\text{nBE}}$.

As an example, we construct the TBE potential.  When the ladder Bethe-Salpeter
equation is used, only one Feynman graph contributes, the box diagram arising
from the iteration of the Feynman OBE kernel.  This gives six non-vanishing
LFTOPT diagrams.  (The other diagrams vanish because the vacuum is simple on
the light front.)
\begin{eqnarray}
\left( \, \FeynLaddergraph{30}{20} \, \right)_{\mbox{Feynman}}
   &=& \, \itOPEgraphOne{30}{20}
\,  +  \, \itOPEgraphTwo{30}{20}
\,  +  \, \itOPEgraphThree{30}{20}
\,  +  \, \itOPEgraphFour{30}{20}
\,  +  \, \sbgraphOne{30}{20}
\,  +  \, \sbgraphTwo{30}{20} \, .
\end{eqnarray}
The first four diagrams are iterations of the OBE potential and are 
reducible with respect to $G_{\text{LF}}$.  The last two are
two-particle-irreducible and thus constitute the TBE potential,
$V_{\text{TBE}}$.

A truncation must be made for the potential in Eq.\ (\ref{fullpot}), since 
in this Hamiltonian theory the infinite sum of graphs (required to reproduce 
the covariant results of the Bethe-Salpeter equation) cannot be calculated.
In this paper we consider two truncations, one where we only keep the OBE
potential, and another where we keep OBE and TBE potentials.  We write out the
matrix elements of $V_{\text{OBE}}$ and $V_{\text{TBE}}$ in the two-particle
momentum basis, denoting the momentum of the incoming particles by $p_1$ and
$p_2$, and the outgoing particles $k_1$ and $k_2$.  For $V_{\text{TBE}}$,
particles 1 and 2 have intermediate momenta which we denote by $q_1$ and
$q_2$.  In terms of diagrams, the potentials are 
\begin{eqnarray}
& & \nonumber \\
V_{\text{OBE}} 
&=& \qquad
\opegraphLOne{45}{30}{{k_2}{p_2}}{{k_1}{p_1}}{p_1-k_1}
\qquad + \qquad 
\opegraphLTwo{45}{30}{{k_2}{p_2}}{{k_1}{p_1}}{k_1-p_1} \label{obegraph} \\
& & \nonumber \\
& & \nonumber \\
& & \nonumber \\
V_{\text{TBE}} 
&=& \qquad
\sbgraphLOne{45}{30}{{k_2}{q_2}{p_2}}{{k_1}{q_1}{p_1}}{q_1-k_1}{p_1-q_1}
\qquad \quad + \quad \qquad
\sbgraphLTwo{45}{30}{{k_2}{q_2}{p_2}}{{k_1}{q_1}{p_1}}{q_1-p_1}{k_1-q_1}
\qquad \quad. \label{tbegraph} \\
& & \nonumber
\end{eqnarray}
These potentials are implicitly functions of the incoming and outgoing 
momenta.

The wavefunction in Eq.\ (\ref{fullse}) can be written as a function of the
momenta of the two particles.  However, since the total $\bbox{P}_\perp$ and
$P^+$ momenta commute with the interaction, the total $\bbox{P}_\perp$ and
$P^+$ of the bound state must be equal to the sum of the two particles'
perpendicular and plus momentum both before and after the interaction.
Therefore, we can take the total momentum  of the bound state as an external
parameter, and solve the light-front Schr\"odinger equation for that total
momentum.  Thus, the wavefunction, when parameterized by the total momentum,is
a function of only one momentum variable.  We can take that momentum to be the
momentum of particle 1.  The components of particle 2's momentum are then
$k_2^+ = P^+ - k_1^+$ and
$\bbox{k}_{2,\perp} = \bbox{P}_\perp - \bbox{k}_{1,\perp}$; the minus component
(the light-front energy) is defined by the requirement that the particle is on
mass shell, so $k^-_2 = (M^2+\bbox{k}_{2,\perp}^2)/k_2^+$.

We also define $x \equiv k_1^+/P^+ = x_{Bj}$, where $x_{Bj}$ is the Bjorken
$x$ variable, so that $k_2^+/P^+=1-x$.  Likewise, we write the Bjorken
variables that correspond to the other momenta in the diagrams in Eqs.\ 
(\ref{obegraph}) and (\ref{tbegraph}) as $y \equiv p_1^+/P^+$ and
$z \equiv q_1^+/P^+$.  Also, a meson with arbitrary momentum $q^+$ and
$\bbox{q}_\perp$ (not to be confused with the loop momentum $q_1$), has
light-front energy $\omega^-(q) = (\mu^2+\bbox{q}_\perp^2)/q^+$.

Inspecting the rules for converting a light-front time-ordered diagram into a
formula \cite{Cooke}, we find that each piece of the potential is proportional
to
\begin{eqnarray}
\frac{1}{2(2\pi)^3 \sqrt{x(1-x)y(1-y)}}.
\end{eqnarray}
We will suppress this term from the following potentials.

For further simplification, we say that we are in the bound state's
center-of-momentum frame so that the total four-momentum
$P^\mu = (P^-,P^+,\bbox{P}_\perp) = (E,E,\bbox{0}_\perp)$, where $E$ is the
bound-state energy.  We can do this without loss of generality since the
boosts are kinematic, so we can easily boost the light-front Schr\"odinger
equation to a frame where the bound state has arbitrary momentum.  All further
calculations will be done in the c.m. frame.

Now we use the light-front time-ordered perturbation theory rules to
calculate the potential due to OBE,
\begin{eqnarray}
V_{\text{OBE}}(\bbox{k}_{1,\perp},x,\bbox{p}_{1,\perp},y;E)
&=& \left(\frac{M}{E}\right)^2
\left[
\frac{ \theta(x-y)/|x-y|}{E - p^-_1 - k^-_2 - \omega^-(k_1-p_1)}
\right. \nonumber \\
&& \phantom{\frac{(gM)^2}{E^2} \times \left[\right.} \left. 
+
\frac{ \theta(y-x)/|y-x|}{E - k^-_1 - p^-_2 - \omega^-(p_1-k_1)}
\right] \label{OBEpot}.
\end{eqnarray}
We also calculate the potential due to the TBE,
\begin{eqnarray}
V_{\text{TBE}}(\bbox{k}_{1,\perp},x,\bbox{p}_{1,\perp},y;E) &=&
\left(\frac{M}{E}\right)^4
\int \frac{d^2 q_\perp}{2(2\pi)^3} \left[
\int_y^x \frac{dz}{z(1-z)(z-y)(x-z)}
\right. \nonumber \\ && \phantom{\left(\frac{M}{E}\right)^4 \int
\frac{d^2 q_\perp}{2(2\pi)^3} \left[\right.} \left. \times
\frac{1}{E -q_1^- - k_2^- - \omega^-(k_1-q_1)}
\right. \nonumber \\ && \phantom{\left(\frac{M}{E}\right)^4 \int
\frac{d^2 q_\perp}{2(2\pi)^3} \left[\right.} \left. \times
\frac{1}{E -p_1^- - k_2^- - \omega^-(k_1-q_1) - \omega^-(q_1-p_1)}
\right. \nonumber \\ && \phantom{\left(\frac{M}{E}\right)^4 \int
\frac{d^2 q_\perp}{2(2\pi)^3} \left[\right.} \left. \times
\frac{1}{E -p_1^- - q_2^- - \omega^-(q_1-p_1)} \right]
\nonumber \\ && \phantom{\left(\frac{M}{E}\right)^4 \int
\frac{d^2 q_\perp}{2(2\pi)^3}} 
+ \{ 1\leftrightarrow 2\} \label{TBEpot}.
\end{eqnarray}
The $\{ 1\leftrightarrow 2\}$ means to replace all labels 1 with 2 and vice
versa, as well as replacing the Bjorken variables $x$, $y$, and $z$ with
$1-x$, $1-y$, and $1-z$.  The angular part the $d^2q_\perp$ integral in Eq.\ 
(\ref{TBEpot}) can be done analytically.  The radial part of that integral 
and the $z$ integral need to be done numerically.  A detailed discussion of
the evaluation of Eq.\ (\ref{TBEpot}) will be given in Ref.\ \cite{Cooke}.

We will find it useful to convert from our light-front coordinates
$(k_1^+,\bbox{k}_\perp)$ to equal-time coordinates
$\bbox{k}=( \bbox{k}_\perp,k^3)$ by using an implicit definition of $k^3$
\cite{Terentev:1976jk}
\begin{eqnarray}
k_1^+ &=& \frac{E}{2E(\bbox{k})} \left[ E(\bbox{k}) + k^3\ \right] \label{eteq}
\\
E(\bbox{k}) &=& \sqrt{ M^2 + \bbox{k}^2 }.
\end{eqnarray}
We will refer to the equal-time vector $\bbox{k}$ as the relative
momentum of the two-particle system.  It is worth emphasizing that we
call $\bbox{k}$ an equal-time vector, however no simple change of
variables can produce an equal-time wavefunction from a light-front
wavefunction.  This is simply a useful change of variables.  The physics
of the light-front remains from the definition of the Hamiltonian
($P^-$) and the commutation relations.

For scattering states, the OBE potential computed for on-shell nucleons and
used in the Weinberg integral equation \cite{Weinberg:1966jm} (which is
essentially the scattering analogue of Eq.\ (\ref{fullse})) leads to
manifestly rotationally
invariant results when written in terms of the relative momenta
\cite{Miller:1997cr}.  The similarity between that rotationally invariant 
result and the usual equal-time result implies that this equal-time momentum
$\bbox{k}$ can be interpreted as the relative momentum of the two particles.
For bound states, this exact simplification does not occur, since for
bound states the potential is, of necessity, evaluated off the energy shell
(but on the mass shell).  However, we expect that the OBE potential, written in
terms of the relative momentum, is {\em approximately} spherically symmetric
for lightly-bound states.  Thus, the wavefunctions are approximate
eigenfunctions of the ``relative angular momentum'', where we define
the ``relative angular momentum'' by the operator
$\bbox{L}=\bbox{x}\times\bbox{k}$.
Our ``relative angular momentum operator'' is not the same as the
true orbital angular momentum operator which is obtained from the
Lagrangian via the energy-momentum tensor in a way similar to the
Hamiltonian.

Now consider the exchange of the particle labels 1 and 2.  This causes
\begin{eqnarray}
\bbox{k}_{1,\perp} &\rightarrow& \bbox{k}_{2,\perp} = -\bbox{k}_{1,\perp} \\
k_1^+ &\rightarrow& k_2^+ = E-k_1^+,
\end{eqnarray}
which means that $k^3$ as defined in Eq.\ (\ref{eteq}) transforms as
$k^3 \rightarrow -k^3$ so $\bbox{k} \rightarrow -\bbox{k}$.
Consequently, exchange of particle labels 1 and 2 is the same as parity.

Since the two nucleons are identical except for the particle label, the
effective potential commutes with parity to all orders in $g^2$. Furthermore,
the light-front Hamiltonian is explicitly invariant under rotations about the
three-axis. These considerations allow us to classify the wavefunctions as
having eigenvalues $p$ of parity (${\mathcal P}$) and $m$ of the
three-component of the angular momentum operator ($J_3$).  We label the
wavefunction as $\langle k_1 |\psi_{m,p}\rangle$, where 
\begin{eqnarray}
\langle k_1 |J_3|\psi^n_{m,p}\rangle &=&
\langle k_1 |    \psi^n_{m,p}\rangle m, \\
\langle k_1 | {\mathcal P} |\psi^n_{m,p}\rangle &=&
\langle k_2   |\psi^n_{m,p}\rangle \\
&=&
\langle k_1 |    \psi^n_{m,p}\rangle p.
\end{eqnarray}

With this, we may write the OBE truncation of full uncrossed Hamiltonian as
\begin{eqnarray}
P^-_{\text{OBE}}(g,E) &=& P_0^- + g^2 V_{\text{OBE}}(E),
\end{eqnarray}
which gives OBE light-front Schr\"odinger equation
\begin{eqnarray}
P^-_{\text{OBE}} \Big( g^{n,m,p}_{\text{OBE}}(E),E \Big)
|\psi^{n,\text{OBE}}_{m,p} \rangle &=&
|\psi^{n,\text{OBE}}_{m,p} \rangle E, \label{obeham}
\end{eqnarray}
where $E$ is an arbitrary energy, $|\psi^{n,\text{OBE}}_{m,p} \rangle$ is the 
$n^{\text{th}}$ wavefunction with parity $p$ and $J_3$ quantum number $m$,
and $g^{n,m,p}_{\text{OBE}}(E)$ is the coupling constant which yields that
bound-state wavefunction with $E$ as the bound-state energy.

For the OBE+TBE truncation, we have
\begin{eqnarray}
P^-_{\text{TBE}}(g,E) &=& P_0^- + g^2 V_{\text{OBE}}(E)
+ g^4 V_{\text{TBE}}(E),
\end{eqnarray}
which gives TBE light-front Schr\"odinger equation
\begin{eqnarray}
P^-_{\text{TBE}} \Big( g^{n,m,p}_{\text{TBE}}(E),E \Big)
|\psi^{n,\text{TBE}}_{m,p} \rangle &=&
|\psi^{n,\text{TBE}}_{m,p} \rangle E, \label{tbeham}
\end{eqnarray}
where the quantities here are defined in an analogous way to Eq.\ 
(\ref{obeham}).  By comparing the spectra $g^{n,m,p}_{\text{OBE}}(E)$ and
$g^{n,m,p}_{\text{TBE}}(E)$, we can see what effect adding the TBE potential
to the OBE potential has on the coupling constant for a given bound-state
energy.

\subsection{Comparison}

The solution to the untruncated light-front Schr\"odinger equation in the
uncrossed approximation is equivalent to the solution of the ladder
Bethe-Salpeter equation.  When the full uncrossed Hamiltonian is truncated,
differences will be introduced.  Thus, here we think of the ladder
Bethe-Salpeter equation as the exact theory which the truncated light-front
Schr\"odinger equations approximate.  As more graphs are included in the
truncated light-front Hamiltonian potential, the agreement with the
Bethe-Salpeter equation will obviously be better.  The question we wish to
answer is how well the spectra $g(E)$ for the two different truncations
[Eqs.\ (\ref{obeham}) and (\ref{tbeham})] approximate the spectra for the
``exact'' theory, the BSE results.

The lack of manifest rotational invariance of the truncated Hamiltonian theory
causes a breaking of the degeneracy of the spectra of the truncated light-front
Schr\"odinger equations for different $m$ states --- unlike the case for the
Bethe-Salpeter equation.  The wavefunctions from the Hamiltonian approach are
classified by their dominant angular momentum contribution $l$, so that we can
compare the spectra for different $m$ projections of the same total angular
momentum $l$.  By doing this, we can compare how the degeneracy of the spectra
is broken for the  OBE and the OBE+TBE truncations, and also compare to the
spectra obtained from the ladder Bethe-Salpeter equation.

\section{Results} \label{results}

For our numerical work, we pick the meson mass to be $0.15$ times that of the
nucleon, so $\mu = 0.15 M$.  This is chosen so that our ground state can be
considered a toy model of deuterium, and also to facilitate comparison with
the results of Schoonderwoerd, Bakker, and Karmanov
\cite{Schoonderwoerd:1998pk}.

The technology for doing these bound-state Bethe-Salpeter equation
calculations was developed over 30 years ago
\cite{LevineWright1,LevineWright2,paganamenta}.
Since the eigenstates of the ladder Bethe-Salpeter equation 
are also eigenstates of the total angular momentum, there is exact degeneracy 
in the energies of the different $m$ states for the same angular momentum.
For the range of parameters used in this study, we find that the
numerical errors in $g^2$ are less than 0.5\%.
The numerical errors are largest for the most deeply-bound states
($E\approx 1.85 M$) with the largest coupling constants
($g^2/4\pi \approx 50$).

\begin{figure}[!ht]
\begin{center}
\epsfig{angle=0,width=3.0in,height=2.4in,file=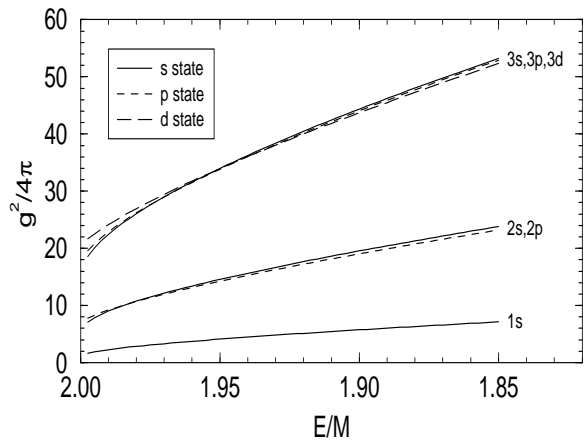}  
\caption{We display here the first three energy bands
for the ladder Bethe-Salpeter equation.
$E$ is the energy of the bound state of two nucleons, and $M$ is the mass
of each of the two nucleons.  The mass of the meson is $\mu = 0.15 M$.
\label{bsfig}}
\end{center}
\end{figure}

The solutions to the ladder Bethe-Salpeter equation form bands where excited
states with different orbital angular momentum are approximately degenerate
with each other, as shown in Fig.\ \ref{bsfig}.  This approximate degeneracy
is due to the fact that when $\mu \rightarrow 0$, this model can be 
shown to have the same degeneracies as the non-relativistic hydrogen atom
\cite{WickCutkosky}.  Thus, in that limit, all states with the same
principal quantum number have the same energy.  When $\mu \neq 0$, that
degeneracy is broken, but only slightly, as we can see from Fig.\ \ref{bsfig}.
Because of this, we will label our states using atomic spectroscopy notation.

Next, we consider the two light-front Schr\"odinger equations given by
Eqs.\ (\ref{obeham}) and (\ref{tbeham}).  These equations are solved
numerically (for each parity and several $m$ values) for the spectrum
$g(E)$ for a range of bound-state energies $E$.  
The symmetries of the light-front Hamiltonian allow us to classify the
states according to $m$ and the action under parity, so as an example,
we calculate the spectra with even parity and $m=0$.
For the range of values we use, we find the numerical errors in $g^2$
are less than 2\%.
The errors are largest for the most deeply bound states with the largest
coupling constants, as was the case for the Bethe-Salpeter equation.

\begin{figure}[!ht]
\begin{center}
\epsfig{angle=0,width=3.0in,height=3.0in,file=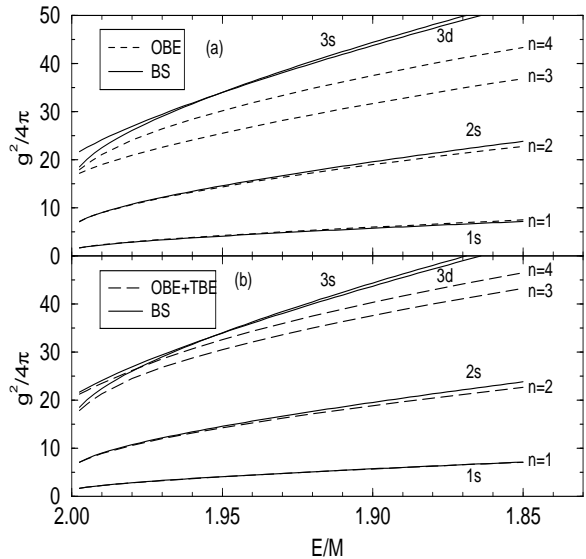}  
\caption{Spectra for the four lowest wavefunctions of even parity and $m=0$
(a) for the OBE equation (short-dashed line) and the Bethe-Salpeter equation
(solid line),
(b) for the OBE+TBE equation (long-dashed line) and the
Bethe-Salpeter equation
(solid line).  The curves are labeled by $n$, which indicates that 
the curve belongs to the $n^{\text{th}}$ eigenvector.
The curve for the $1s$ state for the Bethe-Salpeter equation
and the curve for the first OBE+TBE wavefunction are very close together, but
distinct.
\label{plotraw}}
\end{center}
\end{figure}

We plot both the OBE spectra [in Fig.\ \ref{plotraw}(a)] and the
OBE+TBE spectra [in Fig.\ \ref{plotraw}(b)] along with the
Bethe-Salpeter spectra for the even parity and $m=0$ states.
Based on the energies, we see that
the $n=1$ state is expected to be the $1s$ state,
the $n=2$ state the $2s$ state, and 
the $n=3$ and $n=4$ states the $3s$ or $3d$ states.  We see approximate
agreement between both of the truncated Hamiltonian results and the 
Bethe-Salpeter results for lightly-bound systems (where $E \approx 2M$).

Our states are not manifest eigenstates of total angular momentum $J^2$.
However, if our approximation of the Hamiltonian potential was good
enough, we would be able to identify the $n=3$ and $n=4$ states of the
light-front Hamiltonian calculation with $3s$ and $3d$ states of the
Bethe-Salpeter equation unambiguously.  This would determine the angular
momentum of the light-front Schr\"odinger equation wavefunctions.  From
Fig.\ \ref{plotraw}, we see both that the addition of the TBE potential
brings the $n=3$ and $n=4$ states closer to the Bethe-Salpeter results,
and that in the lightly-bound region the $n=3$ and $n=4$ states can be
identified as $3s$ and $3d$ states.  For more deeply-bound states, this
identification cannot be made.

An alternative approach to assigning angular momenta labels would be
perform a partial-wave decomposition of the wavefunctions in the real
angular momentum basis.  The wavefunctions could then be labeled by the
angular momentum component which is dominant.  In light-front dynamics,
this is difficult to do since the perpendicular components of the real
angular momentum operator ($J_x$ and $J_y$) are dynamical, which makes
the angular momentum operator as complicated as the Hamiltonian.  We are
encouraged to look for an alternative operator which is easier to use,
yet approximates the behavior of the real angular momentum.  We shall
use the ``relative angular momentum'', defined by
$\bbox{L}=\bbox{x}\times\bbox{k}$, where the vectors are equal-time
three-vectors.  The relative momentum $\bbox{k}$ is defined by 
Eq.~(\ref{eteq}) and $\bbox{x}$ is canonically conjugate to $\bbox{k}$.
The ``relative angular momentum'' $\bbox{L}$ is used to help analyze our
solutions, and is not the same as the real angular momentum that can be
derived from the Lagrangian using the energy-momentum tensor.  However,
we will see that the ``relative angular momentum'' gives results that
are expected from the real angular momentum, so our use of the
``relative angular momentum'' in place of the real one appears to be
justified.

A partial-wave decomposition is performed on the wavefunctions (represented
in the relative momentum basis) to obtain the radial wavefunctions
$R_{l,m}$ for all ``relative angular momentum'' states $Y_{l,m}$.
Since the potential is not manifestly rotationally invariant when
written in terms of the relative momentum, our wavefunctions will have support
from many different partial waves.  Not all partial waves are allowed; only
those partial-wave states with the same $J_3$ and parity quantum numbers 
as the wavefunction give non-vanishing radial wavefunctions.  We have
\begin{eqnarray}
\langle \bbox{k} | \psi^n_{m,p} \rangle &=& \sum_{l=m}^\infty 
Y_l^m(\theta,\phi) R^{n,p}_{l,m}(k).
\end{eqnarray}
We define the fraction of the wavefunction with ``relative angular
momentum'' $l$ as
\begin{eqnarray}
f^n_l &=& \int_0^\infty dk\, k^2 \left|R^{n,p}_{l,m}(k)\right|^2.
\end{eqnarray}
The fractions $f^n_l$ are a measure of the amount of ``relative angular
momentum'' state $l$ in the $n^{\text{th}}$ eigenfunction.

\begin{figure}[!ht]
\begin{center}
\epsfig{angle=0,width=3.0in,height=4.0in,file=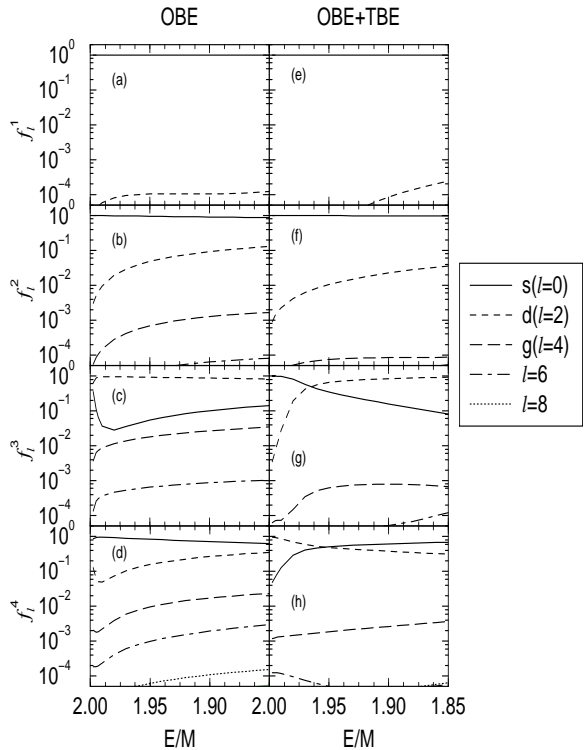}  
\caption{These are the fractions of each angular momentum state in the 
four lowest coupling constant wavefunctions with $m=0$ and even parity
plotted as a function of the binding energy $E/M$.
The results for the OBE truncation are shown on the left, (a-d),
and the OBE+TBE truncation are shown on the right, (e-h).  Note that
the OBE wavefunctions, in general, have more support from a larger number of 
partial waves.
\label{fracfig}}
\end{center}
\end{figure}

To illustrate these ``relative angular momentum'' fractions, we perform
this analysis on the four lowest coupling constant wavefunctions with
even parity and $m=0$ for both the OBE and OBE+TBE truncations, the same
as in Fig.\ \ref{plotraw}.  We show $f^n_l$ as a function of $E$ in
Fig.\ \ref{fracfig}.  (The plots for the $m=\pm1,\pm2$ states show
similar behavior, with different $l$ values.)  Note that more higher
``relative angular momentum'' states contribute for deeper bound
states.  This is the same behavior that would be expected if the real
angular momentum was used to perform the partial wave decomposition.
Recall that the real angular momentum will commute with the full
potential if no truncation is made.  However, the potentials we use are
truncated, and neglecting the higher-order terms breaks the rotational
invariance of the potential.  Also, both the binding and the importance
of the higher-order diagrams increase with the coupling constant.  Thus,
both truncations we consider will break rotational invariance more when
the states are more deeply bound.  In the absence of having the real
angular momentum fractions, we will use the ``relative angular
momentum'' fractions.

Examination of the $f^n_l$ curves in Fig.\ \ref{fracfig} shows us that, as
postulated earlier, the $n=1$ and $n=2$ states shown in Figs.\ 
\ref{fracfig}(a,e) and \ref{fracfig}(b,f) are predominately $s$-wave states,
which we label the $1s$ and $2s$ states respectively.  For the OBE truncation,
the $n=3$ state shown in Fig.\ \ref{fracfig}(c) is predominately $d$-wave 
(labeled the $3d$ state) and the $n=4$ state shown in Fig.\ \ref{fracfig}(d)
is predominately $s$-wave (labeled the $3s$ state), with little mixing between
the two states.  For the OBE+TBE truncation, the $n=3$ and $n=4$ states are
mixtures of both the $3s$ and $3d$ states, as seen in Figs.\ \ref{fracfig}(g,h)
In fact, we see a level crossing between the $n=3,4$ states in that the $n=3$
state is predominantly $3d$ for lightly-bound systems, but as the binding
increases, the $n=3$ state becomes predominantly $3s$.  Again, this
type of mixing would also be found if the real angular momentum was used.

If this $3s$-$3d$ mixing is ignored, then Fig.\ \ref{fracfig} shows clearly
that when the TBE is included the amount of ``relative angular momentum'' states mixed in
actually {\em decreases} as compared to the OBE results.  This implies that the
wavefunctions of the OBE+TBE equation are better ``relative angular momentum'' states than
the wavefunctions of the OBE equation.  Thus the TBE potential restores some
rotational invariance to the OBE calculation.

We attempt to classify the wavefunctions as states with definite angular
momentum.  If a wavefunction has the most support from the ``relative
angular momentum''  state $l$, we classify that state as having angular
momentum $l$.  This
procedure designates the first two wavefunctions for both the OBE and the
OBE+TBE cases as $s$-wave states, which is clearly the right thing to do.  For
the third and fourth wavefunctions of the OBE+TBE equation, there are points
where the fraction of the $s$-wave state equals the fraction of the $d$-wave
state.  Here it no longer clear that such a state should be assigned a
definite ``relative angular momentum'' value; we do so regardless and analyze the
consequences later.

\begin{figure}[!ht]
\begin{center}
\epsfig{angle=0,width=3.0in,height=2.4in,file=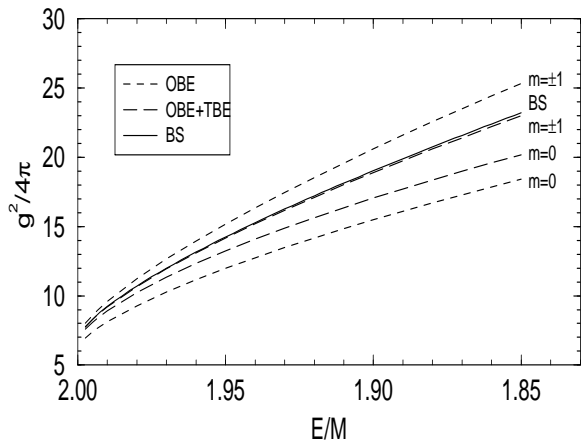}  
\caption{Here we plot the curves for the first $p$-wave state, the $2p$ state.
The Bethe-Salpeter (BS) result is plotted with a solid line, the
OBE results with the short-dashed lines, and the OBE+TBE with the
long-dashed lines.
\label{2pfig}}
\end{center}
\end{figure}

\begin{figure}[!ht]
\begin{center}
\epsfig{angle=0,width=3.0in,height=2.4in,file=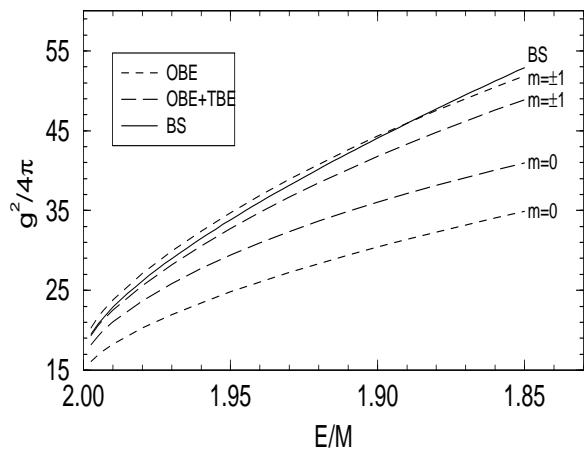}  
\caption{Here we plot the curves for the second $p$-wave state, the $3p$ state.
The Bethe-Salpeter (BS) result is plotted with a solid line, the
OBE results with the short-dashed lines, and the OBE+TBE with the
long-dashed lines.
\label{3pfig}}
\end{center}
\end{figure}

\begin{figure}[!ht]
\begin{center}
\epsfig{angle=0,width=3.0in,height=4.0in,file=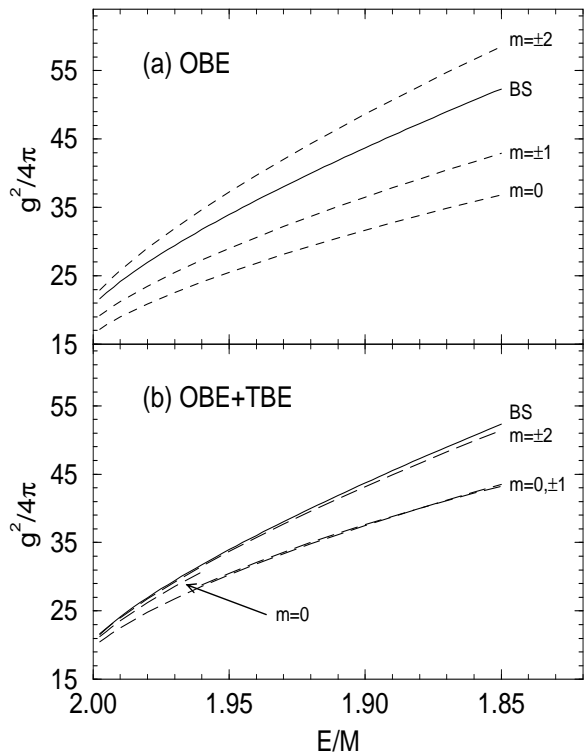}  
\caption{
The first $d$-wave state, the $3d$ state.
The Bethe-Salpeter (BS) result is plotted with a solid line.
In (a) the
OBE results are the short-dashed lines, and in (b) 
the OBE+TBE results are the long-dashed lines.
Note that there are two separate $m=0$ curves when the OBE+TBE
are used, one extending a quarter of the way over, and the other hiding under
the $m=\pm1$ curve with the OBE+TBE.
\label{3dfig}}
\end{center}
\end{figure}

We now examine the breaking of rotational invariance of the two truncations
based on the comparison of states with different $m$ projections of the same
angular momentum.  Since the $s$-wave state has only $m=0$, there is
nothing to analyse in that case.  Further discussion of the ground-state
$s$-wave will appear in Ref.\ \cite{Cooke}.
In Figs.\ \ref{2pfig}, \ref{3pfig}, and \ref{3dfig}, we plot the
Bethe-Salpeter bound-state spectra along with the spectra for the states
constructed with the OBE and the OBE+TBE potentials.  The different $m$ states
for the Bethe-Salpeter equation are exactly degenerate as a result of the
rotational invariance of the equation.  
The curves from the Hamiltonian theory do not exhibit the exact degeneracy
in $m$; the degeneracy is broken whether the OBE or OBE+TBE potentials are 
used.  However, we see that the degeneracy is partially restored when the
TBE potential is included, in that for a given binding energy the spread of
the coupling constants is always smaller.  

In Figs.\ \ref{2pfig} and \ref{3pfig}, we plot the spectra for the $2p$ and
$3p$ states, respectively.  In both figures, we see that the $m=0$ and $m=\pm1$
curves move closer together after addition of the TBE potential.  However, the
average of the two curves for each case does not move much.  In fact, for the
$3p$ state in Fig.\ \ref{3pfig}, the $m=\pm1$ curve moves farther away from the
ladder Bethe-Salpeter curve after addition of the TBE.  We also note that in
Fig.\ \ref{3pfig} the spread of the spectra is larger than in Fig.\ \ref{2pfig}
for both the OBE and the OBE+TBE light-front Schr\"odinger equations.  We
attribute this to the neglect of the higher-order graphs.  Since the states
with different $m$ values would be degenerate if all the higher-order graphs
were included, not including them causes a breaking of the degeneracy.  Because
the coupling constant is larger for the $3p$ states than for the $2p$ states
with the same binding energy, the omission of the higher-order graphs causes a
larger breaking of the degeneracy for the $3p$ states.

We also consider the first $d$-wave states.  We see in Fig.\ \ref{3dfig} that,
as in Figs.\ \ref{2pfig} and \ref{3pfig}, 
the states with different $m$ values move together after addition of TBE.
The effect of the level crossing of the $3s$ and $3d$ states shown in
Figs.\ \ref{fracfig}(g) and \ref{fracfig}(h) is seen here in that there are
two disjoint curves for $m=0$.  When the level crossing occurs, at
$E/M \approx 1.96$, the wavefunction which originally was the $3s$ state
becomes the $3d$ state and vice versa.

There are some problems that this level crossing causes.  In the intermediate
region where the $3s$ and $3d$ states are about equally mixed, it is probably
not physically sensible to call the state a $3s$ or $3d$ state.
However, plotting
the curves gives us an indication of what the wavefunctions are doing, and for
this case we see that the $m=0$ curve is always bounded by the $m=\pm1$ and
$m=\pm2$ curves.  A more restrictive classification scheme would leave a gap
(in $E/M$) between the two $m=0$ curves and it would not be so clear that the
bounding of the $m=0$ curve which we observe here occurs.

Finally, we note that the spread of the curves in the $3d$ case is
approximately the same as the spread in the $3p$ case for each truncation.
This tells us that the higher-order graphs have the same qualitative effect in
both states.

\section{Conclusions} \label{conclusions}

In this paper, we considered two truncations of the light-front Hamiltonian,
the OBE and OBE+TBE truncations.  Using these truncations, a ``relative
angular momentum'' operator was used to study the
partial-wave decompositions of the bound-state wavefunctions.  
We found that the ``relative angular momentum'' operator acting on those
states yield behavior similar to that expected from the real angular
momentum.
This result encourages the use of this operator to classify the states
according to their angular momentum values $l$, and to study the
degeneracy of the spectra for states with the same $l$ value
but with different projections $m$.
We found less breaking of the degeneracy when the
OBE+TBE potential was used than when the OBE potential was used.  Both of
these findings indicate that the OBE+TBE truncation of the Hamiltonian breaks
rotational invariance less than the OBE truncation alone.

However, there is still some discrepancy between our truncated Hamiltonian
spectra and the ladder Bethe-Salpeter spectra.  In general, our OBE+TBE
results for the $p$- and $d$-wave states show deeper binding than the
Bethe-Salpeter results.  Not surprisingly, this disagreement is largest for
the most deeply bound-states where the coupling is largest.  This difference
would be removed if all the higher-order pieces of the potential were
included.  These higher-order pieces are also needed for the full
restoration of rotational invariance.

\acknowledgements

We acknowledge a useful discussion with Vladimir Karmanov about angular
momentum and rotational invariance on the light front.  This work is
supported in part by the U.S.\ Dept.\ of Energy under Grant No.\ 
DE-FG03-97ER4014. 

\appendix

\end{fmffile}
\end{document}